\documentclass[amsmath,amssymb,showpacs]{revtex4}

\usepackage{graphicx}
\usepackage{dcolumn}
\usepackage{bm}
\newcommand{\be}{\begin{equation}}
\newcommand{\ee}{\end{equation}}
\newcommand{\ba}{\begin{eqnarray}}
\newcommand{\ea}{\end{eqnarray}}
\newcommand{\baa}{\begin{eqnarray*}}
\newcommand{\eaa}{\end{eqnarray*}}

\begin{document}
   
\title{Generalized-Ensemble Algorithms for the Isobaric-Isothermal Ensemble}
\author{%
Yoshiharu Mori$^{1}$ and 
Yuko Okamoto$^{1,2}$
}
\affiliation{$^{1}$Department of Physics, Nagoya University \\ Nagoya, Aichi 464-8602, Japan \\
$^{2}$Structural Biology Research Center, Nagoya University \\ Nagoya, Aichi 464-8602, Japan}

\begin{abstract}
We present generalized-ensemble algorithms for isobaric-isothermal 
molecular simulations. In addition to the multibaric-multithermal algorithm and 
replica-exchange method for the isobaric-isothermal ensemble, which have already 
been proposed, we propose a simulated tempering method for this ensemble. 
We performed molecular dynamics simulations with these algorithms for an alanine 
dipeptide system in explicit water molecules to test the effectiveness of the algorithms. 
We found that these generalized-ensemble algorithms are all useful for conformational 
sampling of biomolecular systems in the isobaric-isothermal ensemble.
\end{abstract}

\maketitle

Monte Carlo (MC) and molecular dynamics (MD) simulations with generalized-ensemble algorithms have been widely used for studies of proteins and peptides in order to have efficient conformational sampling (for a review, see, e.g., ref. 1).
Three well-known generalized-ensemble algorithms are the multicanonical algorithm (MUCA)~\cite{berg91,berg92} (the MD version was developed in refs. 4 and 5), replica-exchange method (REM)~\cite{hukushima96} (the method is also referred to as parallel tempering~\cite{marinari98} and the MD version, which is referred to as REMD, was developed in ref. 8), and simulated tempering (ST)~\cite{Lyubartsev92,marinari92}. 
These methods can be used for obtaining accurate physical quantities in the canonical ensemble.
Among them, REM (particularly REMD) is often used because the weight factor is \textit{a priori} known (i.e., the Boltzmann factor), while those for MUCA and ST have to be determined before the simulations.

Multidimensional (or multivariable) extensions of the original generalized-ensemble algorithms have been developed in many ways (see the references in ref. 1) and recently the general formulations have been given in refs. 11-13.

In this letter, we present a multidimensional simulated tempering algorithm in the isobaric-isothermal ($NPT$) ensemble.
Several generalized-ensemble algorithms for the $NPT$ ensemble have been proposed (for example, the multibaric-multithermal (MUBATH) algorithm~\cite{okumura04a,okumura04b,okumura04c,okumura06} and REM in temperature and pressure~\cite{nishikawa00,okabe01,sugita02,paschek04}) and here, we present an ST algorithm for the $NPT$ ensemble to complete generalized-ensemble algorithms for this ensemble.

We first briefly review the generalized-ensemble algorithms for isobaric-isothermal molecular simulations.
Let us consider a physical system that consists of $N$ atoms and that is in a box of a finite volume $V$.
The states of the system are specified by coordinates $r\equiv \{\bm{r}_1,\bm{r}_2,\cdots, \bm{r}_N\}$ and momenta $p\equiv \{\bm{p}_1,\bm{p}_2,\cdots,\bm{p}_N\}$ of the atoms and volume $V$ of the box.
The potential energy $E(r,V)$ for the system is a function of $r$ and $V$.

We first describe MC simulation algorithms for MUCA, REM, and ST in the $NPT$ ensemble.
In these cases, momenta of atoms do not have to be considered.
To make a system an equilibrium state, the detailed balance condition is imposed and a transition probability $w(X\to X')$ from an old state $X$ to a new state $X'$ can be given by the Metropolis criterion.~\cite{metropolis53}

In MUBATH simulations, we introduce a function $\mathcal{H}(E,V)$ and use a weight factor $W_{\text{mbt}}(E,V) \equiv \exp \left[-\beta_0 \mathcal{H}(E,V)\right]$ so that the distribution function $f_{\text{mbt}}(E,V)$ of $E$ and $V$ may be uniform: 
\begin{equation}
	f_{\text{mbt}}(E,V) \propto n(E,V)W_{\text{mbt}}(E,V) = \text{constant},
	\label{eq:dis:mbt}
\end{equation}
where $\beta_0$ is an arbitrary inverse reference temperature defined as $\beta_0 = 1 / k_{\text{B}}T_0$ ($k_{\text{B}}$ is the Boltzmann constant) and $n(E,V)$ is the density of states.

To perform MUBATH MC simulations, the trial moves are generated in the same way as in the usual constant $NPT$ MC simulations~\cite{mcdonald72} and the transition probability from $X \equiv \{s,V\}$ to $X' \equiv \{s',V'\}$ is given by~\cite{okumura04a,okumura04b}
\begin{equation}
	w_{\text{mbt}}(X \to X') = \min \left[ 1, \exp (-\varDelta _{\text{mbt}} ) \right],
\end{equation}
where
\begin{equation}
	\varDelta _{\text{mbt}} = \beta_0 \left\{ \mathcal{H}\left[ E(s',V'),V'\right] - \mathcal{H}\left[ E(s,V),V\right] - Nk_{\text{B}}T_0 \ln (V'/V) \right\} ,
\end{equation}
and $s= \{ \bm{s}_1,\bm{s}_2,\cdots,\bm{s}_N\}$ is the scaled coordinates defined by $\bm{s}_i = V^{-1/3}\bm{r}_i \ (i=1,2,\cdots,N)$.
Here, we are assuming the box is a cube of side $V^{-1/3}$.

In REM simulations, we prepare a system that consists of $M_T \times M_P$ non-interacting replicas of the original system, where $M_T$ and $M_P$ are the number of temperature and pressure values used in the simulation, respectively. 
The replicas are specified by labels $i$ $(i=1,2,\cdots ,M_T\times M_P)$, temperature by $m_t$ $(m_t=1,2,\cdots,M_T)$, and pressure by $m_p$ $(m_p=1,2,\cdots,M_P)$.

To perform REM MC (REMC) simulations, we carry out the following two steps alternately: (1) perform a usual constant $NPT$ simulation in each replica at assigned temperature and pressure and (2) try to exchange the replicas.
If the temperature (specified by $m_t$ and $n_t$) and pressure (specified by $m_p$ and $n_p$) between the replicas are exchanged, the transition probability from $X\equiv \{\cdots,(s^{[i]},V^{[i]};T_{m_t},P_{m_p}),\cdots,(s^{[j]},V^{[j]};T_{n_t},P_{n_p} ),\cdots\}$ to $X'\equiv \{\cdots,(s^{[i]},V^{[i]};T_{n_t},P_{n_p}),\cdots,(s^{[j]},V^{[j]};T_{m_t},P_{m_p} ),\cdots\}$ at the trial is given by~\cite{okabe01,sugita02}
\begin{equation}
	w_{\text{rem}}(X \to X') = \min \left[ 1, \exp (-\varDelta _{\text{rem}} ) \right] ,
	\label{rem_transition1}
\end{equation}
where
\begin{equation}
	\varDelta _{\text{rem}} = (\beta_{m_t} - \beta_{n_t})\left[ E(s^{[j]},V^{[j]}) - E(s^{[i]},V^{[i]}) \right] 
	+ (\beta_{m_t} P_{m_p}- \beta_{n_t} P_{n_p})\left( V^{[j]} - V^{[i]} \right) .
	\label{rem_transition2}
\end{equation}

In ST simulations, we introduce a function $g(T,P)$ and use a weight factor $W_{\text{st}}(E,V;T,P) \equiv \exp[-\beta (E+PV) + g(T,P)]$ so that the distribution function $f_{\text{st}}(T,P)$ of $T$ and $P$ may be uniform:
\begin{equation}
	f_{\text{st}}(T,P) \propto \int_0^{\infty}dV\int_V dr \ W_{\text{st}}[E(r,V),V;T,P] = \text{constant}.
	\label{eq:dis:st}
\end{equation}
From eq. (\ref{eq:dis:st}), it is found that $g(T,P)$ is formally given by
\begin{equation}
	g(T,P) = -\ln \left\{ \int_0^{\infty}dV\int_Vdr \exp \left[ -\beta \left( E(r,V) + PV \right) \right] \right\},
\end{equation}
and the function is the dimensionless Gibbs free energy except for a constant. 

To perform ST MC simulations, we carry out the following two steps alternately: (1) perform a usual constant $NPT$ simulation and (2) try to update the temperature and pressure. 
The transition probability from $X\equiv \{ s,V;T,P\}$ to $X'\equiv \{ s,V;T',P'\}$ for this trial is given by 
\begin{equation}
	w_{\text{st}}(X \to X') = \min \left[ 1, \exp (-\varDelta _{\text{st}} ) \right],
	\label{st_transition1}
\end{equation}
where
\begin{equation}
	\varDelta _{\text{st}} = (\beta ' - \beta)E(s,V) + (\beta ' P' - \beta P)V - \left[ g(T',P') - g(T,P)\right] .
	\label{st_transition2}
\end{equation}

For MD simulations with MUCA, REM, and ST in the $NPT$ ensemble, the actual formulations depend on constant temperature and pressure algorithms.
Here, we employ the MD methods with the Martyna-Tobias-Klein (MTK) algorithm,~\cite{martyna94} whose equations of motion follow Nos\'{e}~\cite{nose84a,nose84b} and Hoover~\cite{hoover85} for the thermostat and Andersen~\cite{andersen80} for the barostat.

To perform MUBATH MD simulations, we solve the usual equations of motion for the MTK algorithm except that it is necessary to modify the equations for $\bm{p}_i$ and $p_{\varepsilon}$ as follows: 
\begin{subequations}
\begin{align}
	\frac{d\bm{p}_i}{dt} &= -\frac{\partial \mathcal{H}}{\partial E}\frac{\partial E}{\partial \bm{r}_i} - \left( 1 + \frac{1}{N} \right) \frac{p_{\varepsilon}}{W}\bm{p}_i - \frac{p_{\xi}}{Q}\bm{p}_i ,\\
	\frac{dp_{\varepsilon}}{dt} &= 3V\left( P_{\text{int}} - \frac{\partial \mathcal{H}}{\partial V} \right) + \frac{1}{N} \sum_{i=1}^{N} \frac{\bm{p}_i^2}{m_i} - \frac{p_{\xi}}{Q}p_{\varepsilon},
\end{align} 
\end{subequations}
where
\begin{equation}
	P_{\text{int}} = \frac{1}{3V} \left[ \sum_{i=1}^{N} \frac{\bm{p}_i^2}{m_i} - \frac{\partial \mathcal{H}}{\partial E} \left( \sum_{i=1}^{N} \bm{r}_i \cdot \frac{\partial E}{\partial \bm{r}_i} + 3V \frac{\partial E}{\partial V} \right) \right],
\end{equation}
$p_{\varepsilon}$ is the momentum associated with the logarithm of $V$, $p_{\xi}$ is the momentum of the thermostat, and $W$ and $Q$ are the masses of barostat and thermostat, respectively.

When we perform MD simulations with REM and ST, the momenta should be rescaled if the replicas are exchanged for the temperature in REM~\cite{sugita99} and the temperature is updated in ST~\cite{mitsutake09b}.

In REM MD (REMD) simulations with the MTK algorithm, let an old state be $X\equiv \{\cdots,(r^{[i]},V^{[i]},p^{[i]},p_{\varepsilon}^{[i]},p_{\xi}^{[i]};T_{m_t},P_{m_p}),\cdots,(r^{[j]},V^{[j]},p^{[j]},p_{\varepsilon}^{[j]},p_{\xi}^{[j]};T_{n_t},P_{n_p} ),\cdots\}$ and a new state $X\equiv \{\cdots,(r^{[i]},V^{[i]},p^{[i]\prime},p_{\varepsilon}^{[i]\prime},p_{\xi}^{[i]\prime};T_{n_t},P_{n_p}),\cdots,(r^{[j]},V^{[j]},p^{[j]\prime},p_{\varepsilon}^{[j]\prime},p_{\xi}^{[j]\prime};T_{m_t},P_{m_p} ),\cdots\}$.
When replica exchange is performed, the momenta should be rescaled as follows:~\cite{mori10}
\begin{subequations}
\begin{align}
	p^{[i]\prime} = \sqrt{\frac{T_{n_t}}{T_{m_t}}}p^{[i]}, \quad p^{[j]\prime} = \sqrt{\frac{T_{m_t}}{T_{n_t}}}p^{[j]} \\
	p_{\varepsilon}^{[i]\prime} = \sqrt{\frac{T_{n_t}W_{n}}{T_{m_t}W_{m}}}p_{\varepsilon}^{[i]}, \quad p_{\varepsilon}^{[j]\prime} = \sqrt{\frac{T_{m_t}W_{m}}{T_{n_t}W_{n}}} p_{\varepsilon}^{[j]} \\
	p_{\xi}^{[i]\prime} = \sqrt{\frac{T_{n_t}Q_{n}}{T_{m_t}Q_{m}}}p_{\xi}^{[i]}, \quad p_{\xi}^{[j]\prime} = \sqrt{\frac{T_{m_t}Q_{m}}{T_{n_t}Q_{n}}} p_{\xi}^{[j]} ,
\end{align} 
\end{subequations}
where $W_m$ and $Q_m$ are the masses in the simulation at $T_{m_t}$ and $P_{m_p}$, and $W_n$ and $Q_n$ are the ones at $T_{n_t}$ and $P_{n_p}$.
Then the transition probability (\ref{rem_transition1}) and (\ref{rem_transition2}) can also be used in the REMD simulations.

In ST MD simulations with the MTK algorithm, let an old state be $X\equiv \{ r,V,p,p_{\varepsilon},p_{\xi};T,P\}$ and a new state $X'\equiv \{ r,V,p',p_{\varepsilon}',p_{\xi}';T',P'\}$.
If $p', p_{\varepsilon}'$ and $p_{\xi}'$ are written as
\begin{equation}
	p' = \sqrt{\frac{T'}{T}}p , \quad
	p_{\varepsilon}' = \sqrt{\frac{T'W'}{TW}}p_{\varepsilon} , \quad
	p_{\xi}' = \sqrt{\frac{T'Q'}{TQ}}p_{\xi} ,
\end{equation}
where $W'$ and $Q'$ are the masses in the simulation at $T'$ and $P'$, then it can be shown that the transition probability is given by eqs. (\ref{st_transition1}) and (\ref{st_transition2}).

While the weight factors in REM simulations are \textit{a priori} known, those in MUBATH and ST simulations have to be determined before the simulations. 
We propose two-dimensional extensions of the replica-exchange multicanonical algorithm (REMUCA)~\cite{sugita00} and the replica-exchange simulated tempering (REST)~\cite{mitsutake00} for the isobaric-isothermal ensemble to overcome the difficulty of determining the weight factors.
First, we perform a REM simulation for the isobaric-isothermal ensemble and then calculate the density of states and the dimensionless Gibbs free energy by the multiple-histogram reweighting techniques,~\cite{ferrenberg89} or the weighted histogram analysis method (WHAM),~\cite{kumar92} using the results of the REM simulation.
The WHAM equations for isobaric-isothermal simulations are given by 
\begin{subequations}
\begin{align}
\label{wham1}
	n(E,V) &= \frac{\displaystyle \sum_{i=1}^{M_T} \sum_{j=1}^{M_P}(1+2\tau_{ij})^{-1}N_{ij}(E,V)}{\displaystyle \sum_{i=1}^{M_T} \sum_{j=1}^{M_P}n_{ij}(1+2\tau_{ij})^{-1} \exp \left[ g_{ij}-\beta_i (E+P_jV) \right]} , \\
	\label{wham2}
	\exp \left( -g_{ij}\right) &= \sum_E \sum_V n(E,V) \exp \left[-\beta_i (E+P_jV) \right] ,
\end{align} 
\end{subequations}
where $g_{ij}\equiv g(T_i,P_j)$, $N_{ij}(E,V)$ is the histogram of $E$ and $V$, $n_{ij}$ is the number of the data, and $\tau_{ij}$ is the integrated autocorrelation time in the simulation at $T_i$ and $P_j$, respectively.
These equations need to be solved self-consistently.
We remark that for biomolecular systems, it was found that $\tau_{ij}$ may be set to a constant value for all $i$ and $j$ in eq. (\ref{wham1}).~\cite{kumar92}

In the isobaric-isothermal version of REST, the weight factor can also be determined using the dimensionless Gibbs free energy obtained from eqs. (\ref{wham1}) and (\ref{wham2}).

In order to verify that the generalized-ensemble algorithms discussed above can be effective for conformational sampling and give the same results, we performed MD simulations with the three generalized-ensemble algorithms. 
We used a system of an alanine dipeptide in 73 surrounding water molecules and the system was placed in a cubic cell with periodic boundary conditions.
Both of the backbone dihedral angles $\phi$ and $\psi$ of the peptide were initially set to 180$^{\circ}$.
The simulations were performed with the \textsc{tinker} program package.~\cite{ponder87}
Several of the programs were modified and a few programs were added so that MUBATH, REM, and ST simulations with the MTK algorithm can be performed.
We used the AMBER parm99 force field~\cite{wang00} for the peptide and the TIP3P water model.~\cite{jorgensen83}
The electrostatic interactions were calculated by the particle mesh Ewald method.~\cite{darden93,essmann95}
In the van der Waals interaction calculations, we used the spherical cutoff method and the cutoff distance was set to 12.0 \AA .
The integration of the equations of motion was employed by the method proposed by Martyna \textit{et al.}~\cite{martyna96} and the Shake/Rattle/Roll constraint method~\cite{ryckaert77,andersen83,martyna96} was used so that the water molecules are rigid body molecules.
The unit time step was set to 0.5 fs.
The mass parameters $W$ and $Q$ were determined as in ref. 39.

First, we performed the two-dimensional REMD simulation.
The simulation time was set to 2.0 ns.
We used the following 6 temperature ($T_1,\cdots,T_6$) and 4 pressure ($P_1,\cdots,P_4$) values: 280, 305, 332, 362, 395, and 430 K for temperature and 0.1, 65, 150, and 250 MPa for pressure. 
At the replica-exchange trial, either exchanging temperature ($T$-exchange) or exchanging pressure ($P$-exchange) was chosen randomly and then the pairs $\left\{ (T_1,T_2),(T_3,T_4),(T_5,T_6)\right\}$ or $\left\{ (T_2,T_3),(T_4,T_5)\right\}$ for $T$-exchange and $\left\{ (P_1,P_2),(P_3,P_4)\right\}$ or $\left\{ (P_2,P_3)\right\}$ for $P$-exchange were also chosen randomly.

 
 \begin{figure}
 \begin{center}
 	\includegraphics[width=12cm,clip]{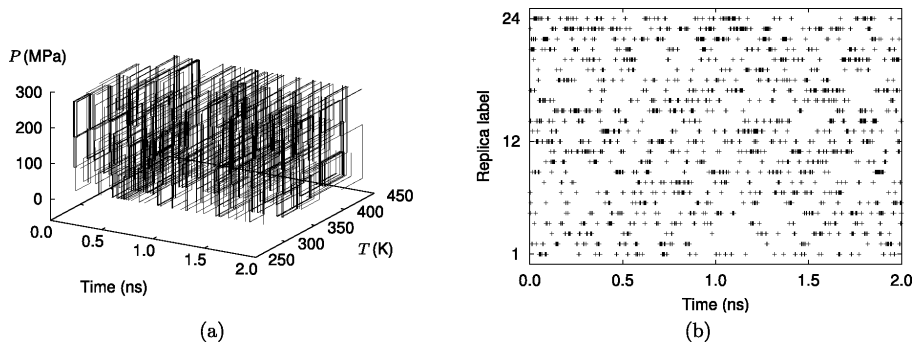}
 \end{center}
 \caption{Results of the two-dimensional REMD simulation: (a) the time series of $T$ and $P$ in one of the replicas and (b) the time series of the replica label at 280 K and 250 MPa.}
 \label{result:remd}
 \end{figure}
 Figure \ref{result:remd}(a) shows the time series of $T$ and $P$ in one of the replicas and Fig. \ref{result:remd}(b) shows the time series of the label of the replicas in the simulation at 280 K and 250 MPa.
 From these figures, it is found that random walks in $T$-$P$ space were realized in each of the replicas.
 The acceptance ratios in the REM simulation were in the range from 0.152 to 0.205 for $T$-exchange and from 0.296 to 0.446 for $P$-exchange, respectively, which also implies that sufficient number of replica exchange was realized.
 
 We then performed 24 MUBATH simulations of 2.0 ns, where the total simulation time was 48 ns so that it is equal to that in the REMD simulation.
 In each of the simulations, different initial velocities were given. 
 The data were sampled every 100 fs.
 The reference temperature was set to 430 K.
 $\partial \mathcal{H}/\partial E$ and $\partial \mathcal{H}/\partial V$ were interpolated by the third-order polynomial, following ref. 17.
 
 \begin{figure}[b]
 \begin{center}
 	\includegraphics[width=12cm,clip]{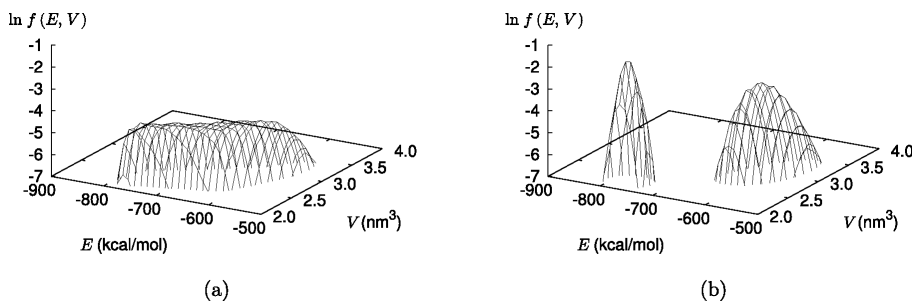}
 \end{center}
 \caption{Logarithm of probability distributions $f$ of $E$ and $V$ (a) in the MUBATH simulations and (b) for the $NPT$ ensemble at 280 K and 250 MPa (left) and at 430 K and 0.1 MPa (right). The distributions corresponding to the $NPT$ ensemble were calculated by the single-histogram reweighting techniques~\cite{ferrenberg88} using the results of the MUBATH simulations.}
 \label{result:mubath}
 \end{figure}
 Figure \ref{result:mubath} shows the probability distributions of $E$ and $V$ from the MUBATH simulations.
 From Fig. \ref{result:mubath}(a), it is found that the MUBATH simulations gave a uniform distribution in the range where the density of states was obtained accurately in the REMD simulation and that the distribution of the MUBATH simulations was much wider than the ones for the $NPT$ ensemble (see Fig. \ref{result:mubath}(b)).
 
 Twenty-four ST simulations of 2.0 ns were carried out, which gave us the same number of sampled data as in the REMD and MUBATH simulations.
 In each simulation, different initial velocities were given. 
 We used the same temperature and pressure values as in the REMD simulation.
 We tried to update the parameter ($T$ or $P$) every 100 fs and the data were sampled just before the trials.
 The choice between updating $T$ ($T$-update) and updating $P$ ($P$-update) was made randomly and then either $T_{m_t-1}$ or $T_{m_t+1}$ for $T$-update and $P_{m_p-1}$ or $P_{m_p+1}$ for $P$-update was also chosen randomly, where $m_t$ and $m_p$ stand for the labels corresponding to the temperature and the pressure before the trial, respectively. 
 
 \begin{figure}[t]
 \begin{center}
 	\includegraphics[width=12cm,clip]{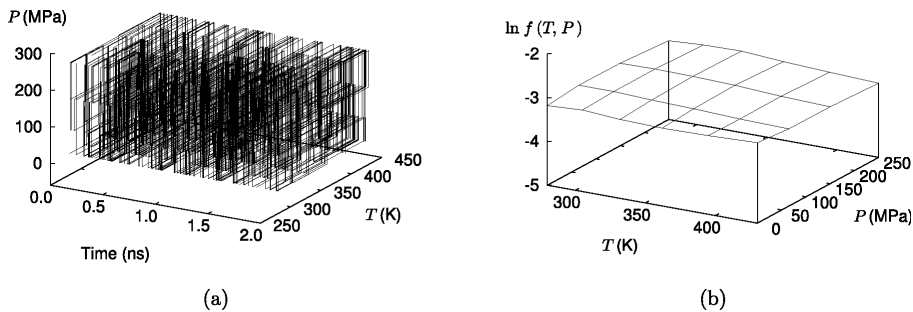}
 \end{center}
 \caption{Results of the ST simulations: (a) the time series of $T$ and $P$ for 2.0 ns and (b) the logarithm of the probability distribution $f$ of $T$ and $P$. }
 \label{result:st}
 \end{figure}
 Figure \ref{result:st}(a) shows the time series of $T$ and $P$ and Fig. \ref{result:st}(b) shows the probability distribution of $T$ and $P$.
 These figures indicate that random walks in $T\mathchar`-P$ space were successfully realized.
 The acceptance ratios in the ST simulations were in the range from 0.260 to 0.430 for $T$-update and from 0.426 to 0.642 for $P$-update, respectively.
 
 \begin{figure}[b]
 \begin{center}
 	\includegraphics[width=8cm,clip]{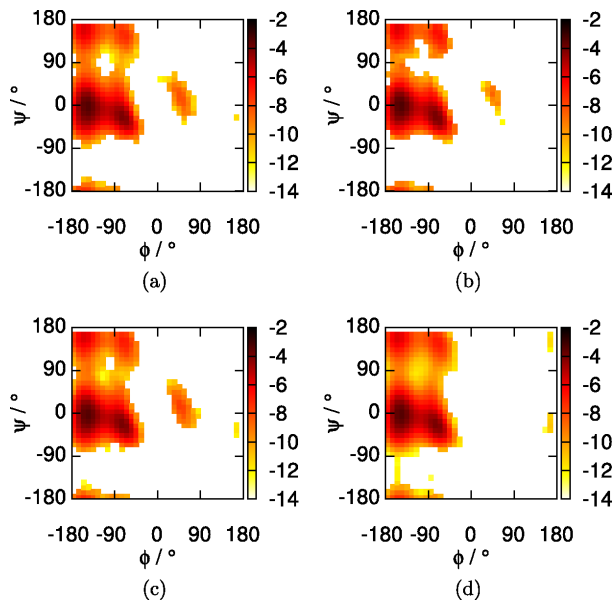}
 \end{center}
 \caption{(Color online) Contour maps of probability distribution of the backbone dihedral angles $\phi$ and $\psi$ in the simulations with (a) REMD, (b) MUBATH, and (c) ST and (d) in the conventional isobaric-isothermal simulations. In these figures, the probability distributions at 298 K and 0.1 MPa are plotted in logarithmic scale and were calculated by the single-histogram reweighting techniques for MUBATH and by WHAM for REMD and ST.}
 \label{fig:dihedral}
 \end{figure}
 Figure \ref{fig:dihedral} shows the probability distributions of the backbone dihedral angles at 298 K and 0.1 MPa in all the simulations.
 Compared with the simulations with REMD, MUBATH, and ST, we also performed 24 conventional isobaric-isothermal simulations of 2.0 ns at 298 K and 0.1 MPa with different initial velocities. 
 The dihedral angle distributions in the simulations with the generalized-ensemble algorithms had a small peak in $0^{\circ}\le \phi \le 90^{\circ}$ and $-90^{\circ} \le \psi \le 90^{\circ}$, although there was no peak in the range in the distribution of the conventional simulations. 
 All the simulations with the three generalized-ensemble algorithms were able to provide the same results and reproduce the distribution obtained previously in refs. 43 and 44.
 
In this letter, we presented the extensions of MUCA, REM, and ST for isobaric-isothermal molecular simulations.
These algorithms can be effective methods for conformational sampling and give accurate physical quantities in the isobaric-isothermal ensemble.
Therefore, one can use the generalized-ensemble algorithms to study temperature and pressure effects on complex systems such as biomolecular systems.

We expect the simulated tempering to be a more useful method for simulations in large systems.
Many replicas are needed in REM simulations of large systems, while only one replica is used in ST simulations.
In addition, ST can be easily implemented into existing program packages compared to MUCA (or MUBATH), in which a force calculation part of programs has to be modified for MD simulations.
Therefore, molecular simulations with ST will be more easily applicable to larger systems than with the other two generalized-ensemble algorithms. 

\section*{Acknowledgments}

Some of the computations were performed on the supercomputers at the Information Technology Center, Nagoya University and at the Research Center for Computational Science, Institute for Molecular Science.
This work was supported, in part, by Grants-in-Aid for Scientific Research on Innovative Areas (``Fluctuations and Biological Functions") and for the Next Generation Super Computing Project, Nanoscience Program from the Ministry of Education, Culture, Sports, Science and Technology (MEXT), Japan.

\end{document}